\documentclass[twocolumn,tighten]{aastex62}
\usepackage{graphicx,hyperref,breakurl,amsmath}

\shorttitle{Fe-Peak Abundances in Dwarf Galaxies}
\shortauthors{Kirby et al.}
\begin{document}

\newcommand{\crdiff}{+0.11}
\newcommand{\crdifferr}{0.03}
\newcommand{\mcr}{+0.58}
\newcommand{\sdcr}{1.16}
\newcommand{\mcrerr}{0.16}
\newcommand{\sdcrerr}{0.11}
\newcommand{\codiff}{-0.01}
\newcommand{\codifferr}{0.04}
\newcommand{\mco}{-0.01}
\newcommand{\sdco}{0.95}
\newcommand{\mcoerr}{0.15}
\newcommand{\sdcoerr}{0.11}
\newcommand{\nidiff}{+0.01}
\newcommand{\nidifferr}{0.01}
\newcommand{\mni}{-0.10}
\newcommand{\sdni}{1.43}
\newcommand{\mnierr}{0.14}
\newcommand{\sdnierr}{0.10}
\newcommand{\nhrsgc}{59}
\newcommand{\nhrshalo}{9}
\newcommand{\nhrsdsph}{60}
\newcommand{\nhrs}{128}
\newcommand{\nhrsrefs}{23}

\newcommand{\crsyserr}{0.123}
\newcommand{\crnltesyserr}{0.153}
\newcommand{\cosyserr}{0.169}
\newcommand{\conltesyserr}{0.182}
\newcommand{\nisyserr}{0.077}

\newcommand{\ngc}{2276}
\newcommand{\nhalo}{16}
\newcommand{\ndsph}{1819}
\newcommand{\ntot}{4111}
\newcommand{\medcrfeerr}{0.19}
\newcommand{\medcrfenlteerr}{0.21}
\newcommand{\medcofeerr}{0.21}
\newcommand{\medcofenlteerr}{0.21}
\newcommand{\mednifeerr}{0.12}
\newcommand{\ncrdup}{322}
\newcommand{\ncrnltedup}{319}
\newcommand{\ncodup}{342}
\newcommand{\nconltedup}{304}
\newcommand{\nnidup}{768}
\newcommand{\meancrdup}{0.03}
\newcommand{\meancrnltedup}{0.01}
\newcommand{\meancodup}{0.07}
\newcommand{\meanconltedup}{0.06}
\newcommand{\meannidup}{-0.05}
\newcommand{\meancrduperr}{0.06}
\newcommand{\meancrnlteduperr}{0.06}
\newcommand{\meancoduperr}{0.05}
\newcommand{\meanconlteduperr}{0.06}
\newcommand{\meanniduperr}{0.04}
\newcommand{\sdcrdup}{0.93}
\newcommand{\sdcrnltedup}{0.91}
\newcommand{\sdcodup}{0.69}
\newcommand{\sdconltedup}{0.68}
\newcommand{\sdnidup}{1.02}
\newcommand{\sdcrduperr}{0.04}
\newcommand{\sdcrnlteduperr}{0.04}
\newcommand{\sdcoduperr}{0.03}
\newcommand{\sdconlteduperr}{0.03}
\newcommand{\sdniduperr}{0.03}

\title{Catalog of Chromium, Cobalt, and Nickel Abundances in Globular
  Clusters and Dwarf Galaxies\footnote{The data presented herein were
    obtained at the W.~M.~Keck Observatory, which is operated as a
    scientific partnership among the California Institute of
    Technology, the University of California and the National
    Aeronautics and Space Administration. The Observatory was made
    possible by the generous financial support of the W.~M.~Keck
    Foundation.}}

\correspondingauthor{Evan N. Kirby}
\email{enk@astro.caltech.edu}

\author[0000-0001-6196-5162]{Evan N. Kirby}
\affiliation{California Institute of Technology, 1200 E.\ California Blvd., MC 249-17, Pasadena, CA 91125, USA}

\author{Justin L. Xie}
\affiliation{California Institute of Technology, 1200 E.\ California Blvd., MC 249-17, Pasadena, CA 91125, USA}
\affiliation{The Harker School, 500 Saratoga Ave., San Jose, CA 95129}

\author{Rachel Guo}
\affiliation{California Institute of Technology, 1200 E.\ California Blvd., MC 249-17, Pasadena, CA 91125, USA}
\affiliation{Irvington High School, 41800 Blacow Rd, Fremont, CA 94538}

\author{Mikhail Kovalev}
\affiliation{Max-Planck Institute for Astronomy, D-69117, Heidelberg, Germany}

\author{Maria Bergemann}
\affiliation{Max-Planck Institute for Astronomy, D-69117, Heidelberg, Germany}


\begin{abstract}

  
We present measurements of the abundances of chromium, cobalt, and
nickel in \ntot\ red giants, including \ngc\ stars in globular
clusters, \ndsph\ stars in the Milky Way's dwarf satellite galaxies,
and \nhalo\ field stars.  We measured the abundances from mostly
archival Keck/DEIMOS medium-resolution spectroscopy with a resolving
power of $R \sim 6500$ and a wavelength range of approximately
6500--9000~\AA\@.  The abundances were determined by fitting spectral
regions that contain absorption lines of the elements under
consideration.  We used estimates of temperature, surface gravity, and
metallicity that we previously determined from the same spectra.  We
estimated systematic error by examining the dispersion of abundances
within mono-metallic globular clusters.  The median uncertainties for
[Cr/Fe], [Co/Fe], and [Ni/Fe] are \medcrfeerr, \medcofeerr,
and \mednifeerr, respectively.  Finally, we validated our estimations
of uncertainty through duplicate measurements, and we evaluated the
accuracy and precision of our measurements through comparison to
high-resolution spectroscopic measurements of the same stars.

\end{abstract}

\keywords{galaxies: dwarf --- galaxies: abundances --- Local Group --- nuclear reactions, nucleosynthesis, abundances --- supernovae}


\section{Introduction}
\label{sec:intro}

The iron group consists of elements close to iron ($Z=26$) in the
periodic table, and it is generally accepted to encompass the elements
from chromium ($Z=24$) to nickel ($Z=28$).  These elements form a
local maximum in the cosmic abundance distribution, peaking at iron.
The root cause of the local maximum is that these elements have high
binding energies per nucleon relative to the adjacent elements.  As a
result, it is energetically favorable to create these elements in
environments with sufficiently high temperature and density.

Different types of fusion create elements from different parts of the
periodic table.  Whereas hydrostatic burning generates most of the
lighter elements, explosive burning creates elements in the iron peak
\citep[i.a.,][]{woo95}.  Therefore, supernovae are the sites of
iron-peak nucleosynthesis.  In fact, thermonuclear (Type~Ia)
supernovae are so efficient at generating iron-peak elements that
their optical light curves are dominated by the radioactive decay of
nickel and cobalt \citep{arn69}.  Type~Ia supernovae generate large
amounts of ${}^{56}$Ni, which is the nuclide with the highest binding
energy per nucleon.  That nuclide radioactively decays into
${}^{56}$Co, which in turn decays into ${}^{56}$Fe.

Abundances of iron-peak elements can probe the physics of explosive
nucleosynthesis.  For example, the thermonuclear explosions of massive
white dwarfs would be expected to generate relatively large amounts of
neutron-rich nuclides compared to the explosions of less massive white
dwarfs.  The reason is that massive white dwarfs are very dense.  High
densities, coupled with electron degeneracy, favor the production of
neutrons.  Therefore, white dwarfs that explode at or near the
Chandrasekhar mass would be expected to produce plenty of neutron-rich
nuclides, including stable manganese \citep[e.g.,][]{sei13}, which has
only one stable isotope: ${}^{55}$Mn.  This isotope is quite
neutron-rich.  Therefore, detecting large amounts of Mn in long-lived
stars might indicate that the Type~Ia supernovae that contributed to
the abundances of these stars exploded at or near the Chandrasekhar
mass.

This article addresses the abundances of iron-peak elements in and
around the Milky Way.  We measure these abundances in mostly archival
spectra of red giants, described in Section~\ref{sec:obs}.  Nearly all
of the giants in our sample are members of dwarf satellite galaxies or
globular clusters (GCs).  As described in Section~\ref{sec:measure},
our innovation is to use medium-resolution spectroscopy rather than
high-resolution spectroscopy, which allows us to analyze a sample size
of thousands of stars.  Because our measurements are not based on a
traditional equivalent width analysis, we devote some time to the
estimation of uncertainties (Section~\ref{sec:errors}) and to the
validation of our measurements through duplicate measurements and
comparison to high-resolution spectroscopy in the literature
(Section~\ref{sec:validation}).  Section~\ref{sec:summary} summarizes
this article.


\section{Observations}
\label{sec:obs}

We use medium-resolution spectroscopy to measure abundances of
iron-peak elements in stars in dwarf galaxies, GCs, and in the field
of the Milky Way halo.  The spectra come from the Keck Deep Imaging
Multi-Object Spectrograph \citep[DEIMOS,][]{fab03}.  The vast majority
of the data in this paper is archival, but we also obtained a few new
spectra.

\subsection{Archival Spectra}
\label{sec:archival}

We rely primarily on the DEIMOS spectra from \citet[][henceforth
  called K10]{kir10}.  K10 observed 2961 red giants in eight dwarf
galaxies, 445 red giants in twelve GCs, and 21 field giants in the
Milky Way halo.  Our sample also includes eight dwarf galaxies
observed by \citet{sim07}.  \citet{kir08b,kir13b} used those spectra
to measure atmospheric parameters like temperature and metallicity.
We also include the spectra of 2227 red giants in 25 GCs observed by
\citet[][henceforth called K16]{kir16}.  Some of those spectra overlap
with K10's sample.

All of the spectra were obtained with DEIMOS via custom slitmasks for
the dwarf galaxies and GCs and via a longslit mask for the halo stars.
DEIMOS was configured with the 1200G grating positioned with a central
wavelength of 7800~\AA\@.  The grating has a groove spacing of
1200~mm$^{-1}$ and a blaze wavelength of 7760~\AA\@.  It provides a
roughly constant dispersion of 1.2~\AA, which corresponds to a
resolving power of $R \sim 6500$ at the central wavelength.  The
wavelength range was approximately 6500--9000~\AA, but the exact range
varied from spectrum to spectrum depending on the location of the slit
within the slitmask.  The OG550 filter was used to block light from
the second and higher diffraction orders.

Flat-fielding was accomplished with a Qz lamp projected onto the
telescope dome.  Wavelength calibration was based on exposures of Ne,
Ar, Kr, and Xe arc lamps turned on simultaneously.  The images were
reduced with the spec2d pipeline \citep{coo12,new13}.  The pipeline
provides one-dimensional spectra that are flat-fielded,
wavelength-calibrated, and background-subtracted.

\subsection{New Spectra}
\label{sec:newspec}

\begin{deluxetable*}{llcccccc}
\tablewidth{0pt}
\tablecolumns{8}
\tablecaption{New DEIMOS Observations\label{tab:newspec}}
\tablehead{\colhead{Target} & \colhead{Slitmask} & \colhead{Targets} & \colhead{UT Date} & \colhead{Airmass} & \colhead{Seeing ('')} & \colhead{Exposures} & \colhead{Exp.\ Time (s)}}
\startdata
NGC 4590   & 4590c-1  & \phn 94 & 2016 Jan 29     & 1.5 & \nodata & 3 &     3600 \\
           & 4590c-2  & \phn 94 & 2016 May 11     & 1.5 & 1.2     & 2 &     2340 \\
NGC 5053   & n5053r-1 &     120 & 2016 Jan 29     & 1.5 & \nodata & 3 &     3600 \\
           & n5053r-2 &     120 & 2016 Mar 5\phn  & 1.0 & 0.5     & 3 &     3600 \\
NGC 5272   & 5272c2   &     134 & 2016 Dec 29     & 1.1 & 0.7     & 1 & \phn 520 \\
           & n5272f-1 &     167 & 2016 Jan 29     & 1.2 & \nodata & 3 &     3600 \\
           & n5272f-2 &     167 & 2016 May 11     & 1.1 & 0.7     & 3 &     3600 \\
NGC 5897   & 5897ar-1 &     123 & 2016 Jan 29     & 1.4 & \nodata & 3 &     3540 \\
           & 5897ar-2 &     123 & 2016 May 11     & 1.4 & 1.0     & 3 &     3600 \\
Palomar 14 & pal14a   & \phn 40 & 2011 Aug 6\phn  & 1.2 & 1.2     & 3 &     3960 \\
Sextans    & sexmi1   & \phn 74 & 2010 May 11     & 1.1 & 0.5     & 3 &     3600 \\
           & sexmi4   & \phn 76 & 2010 May 12     & 1.3 & 0.7     & 3 &     3600 \\
Ursa Minor & umima1   &     122 & 2010 May 11     & 1.5 & 0.8     & 3 &     3600 \\
           & umima2   & \phn 81 & 2010 May 11     & 1.5 & 0.6     & 3 &     3600 \\
           & umima3   & \phn 64 & 2010 May 11     & 1.5 & 0.7     & 3 &     3600 \\
           & umimi1   & \phn 87 & 2010 May 12     & 1.5 & 0.6     & 3 &     3600 \\
           & umimi2   & \phn 76 & 2010 May 12     & 1.5 & 0.7     & 3 &     3600 \\
           & umimi3   & \phn 67 & 2010 May 12     & 1.5 & 0.8     & 3 &     3600 \\
           & umimx1   & \phn 83 & 2010 May 11     & 1.6 & 0.8     & 3 &     3600 \\
           & umimx2   & \phn 81 & 2010 May 11--12 & 1.8 & 0.9     & 2 &     2700 \\
           & umimx4   &     135 & 2010 May 12     & 1.6 & 0.8     & 3 &     3600 \\
\enddata
\end{deluxetable*}

We also obtained new spectra with DEIMOS in five GCs as well as two
dwarf galaxies.  Table~\ref{tab:newspec} gives the observing log for
these new spectra.  Most of the new GC data are repeat observations of
the same stars observed by K16\@.  These data were obtained in
preparation for a study on the binarity of stars in GCs (E.N.\ Kirby
et al., in preparation).  We use the spectra in this work simply
because they are available.  Some of these masks were observed on two
different dates.  Table~\ref{tab:newspec} identifies these separate
observations with the suffix 1 or 2\@.  We also obtained new spectra
in Sextans and Ursa Minor.  The slitmasks were placed along the minor
axis of Sextans.  Hence, they are called ``sexmi.''  In Ursa Minor,
the slitmasks were placed along the major and minor axes (``umima''
and ``umimi'') as well as several other locations near the center of
the galaxy (``umix'').

The first four GCs in Table~\ref{tab:newspec} were also observed by
K16\@.  The slitmask design was the same as described previously.  To
reiterate K16, the photometry was taken from
P.B. Stetson,\footnote{Communicated privately and available at
  \url{http://www2.cadc-ccda.hia-iha.nrc-cnrc.gc.ca/community/STETSON/standards/}.
  See K16 for details.} \citet{wal94}, \citet{ste00}, \citet{tes01},
and \citet{an08}.  We designed the slitmask for Palomar~14 from
Stetson's photometry and \citeauthor{sah05}'s (\citeyear{sah05})
photometry.  The dwarf galaxy slitmasks were designed with photometry
from the Sloan Digital Sky Survey \citep{aba09}.  Stars were selected
to have the approximate colors and magnitudes of red giants at the
distances of their respective galaxies.  Theoretical isochrones
\citep{gir04} were employed to help identify candidate stars in the
color--magnitude diagram.

All of the observations were conducted with the same spectrographic
configuration as described in Section~\ref{sec:archival}.  The
two-dimensional spectra from the same night or adjacent nights were
co-added together.  However, spectra taken in different months were
not co-added.  They are listed on separate lines on
Table~\ref{tab:newspec}.

K10 measured metallicities and some detailed abundance ratios of the
stars in the sample described in Section~\ref{sec:archival}, and we
followed the same procedure for the new spectra.  Part of that
procedure involves determining the spectral continuum.  Following the
same procedure as K10, we estimated the S/N per pixel as the median
absolute deviation from the continuum in the spectral regions
classified as continuum regions.  To convert to S/N per \AA, we
multiplied by $\sqrt{0.33}$, where 0.33 is the number of \AA\ per
pixel.

The elements previously measured by K10 are Mg, Si, Ca, Ti, and Fe.
In this work, we add Cr, Co, and Ni to the elemental abundances
measured from these spectra.  Table~\ref{tab:targets} lists the dwarf
galaxies and GCs that make up the sample, along with the number of
individual stars in each system with measurements of each of those
three elements.  Only stars with estimated uncertainties of less than
0.3~dex are included.

\startlongtable
\begin{deluxetable}{lrrrr}
\tablewidth{0pt}
\tablecolumns{5}
\tablecaption{Spectroscopic Targets\label{tab:targets}}
\tablehead{\colhead{Target} & \colhead{$N({\rm Cr})$} & \colhead{$N({\rm Co})$} & \colhead{$N({\rm Ni})$} & \colhead{$N(\rm any)$}}
\startdata
\cutinhead{Dwarf Galaxies}
Sculptor          &  136 &  162 &  310 &  311 \\
Fornax            &  149 &  301 &  397 &  401 \\
Ursa Major II     &    0 &    1 &    2 &    3 \\
Leo T             &    0 &    0 &    4 &    4 \\
Leo I             &  183 &  310 &  560 &  560 \\
Sextans           &   19 &   12 &   45 &   45 \\
Ursa Major I      &    1 &    1 &    7 &    7 \\
Leo II            &   47 &   69 &  152 &  155 \\
Leo IV            &    0 &    0 &    1 &    1 \\
Coma Berenices    &    1 &    2 &    5 &    5 \\
Canes Venatici II &    0 &    1 &    0 &    1 \\
Canes Venatici I  &   12 &   13 &   50 &   53 \\
Ursa Minor        &   50 &   32 &  143 &  146 \\
Hercules          &    2 &    1 &    4 &    4 \\
Draco             &   48 &   44 &  121 &  123 \\
\cutinhead{Globular Clusters}
NGC 288           &   88 &   88 &  114 &  114 \\
Palomar 2         &    2 &    5 &    9 &   10 \\
NGC 1904 (M79)    &   95 &   86 &  136 &  136 \\
NGC 2419          &   43 &   44 &   74 &   74 \\
NGC 4590 (M68)    &   25 &   53 &   96 &   99 \\
NGC 5024 (M53)    &   16 &   19 &   41 &   41 \\
NGC 5053          &   11 &    9 &   29 &   31 \\
NGC 5272 (M3)     &   54 &   48 &   78 &   78 \\
NGC 5634          &   34 &   37 &   65 &   65 \\
NGC 5897          &   62 &   71 &  206 &  208 \\
NGC 5904 (M5)     &   53 &   50 &   56 &   56 \\
Palomar 14        &    1 &    3 &    5 &    5 \\
NGC 6205 (M13)    &   48 &   36 &   69 &   69 \\
NGC 6229          &   11 &   10 &   18 &   18 \\
NGC 6341 (M92)    &   33 &   42 &  115 &  125 \\
NGC 6656 (M22)    &   44 &   42 &   48 &   49 \\
NGC 6779 (M56)    &   35 &   29 &   57 &   57 \\
NGC 6838 (M71)    &   35 &   42 &   46 &   46 \\
NGC 6864 (M75)    &   73 &   69 &  126 &  127 \\
NGC 7006          &   32 &   32 &   47 &   47 \\
NGC 7078 (M15)    &   99 &  157 &  294 &  300 \\
NGC 7089 (M2)     &  258 &  253 &  363 &  367 \\
NGC 7099 (M30)    &   29 &   43 &  123 &  125 \\
Palomar 13        &    2 &    1 &    8 &    8 \\
NGC 7492          &   10 &    6 &   21 &   21 \\
\cutinhead{Field Stars}
Milky Way halo    &    8 &    7 &   15 &   16 \\
\cutinhead{Total}
Total             & 1849 & 2231 & 4060 & 4111 \\
\enddata
\end{deluxetable}

\subsection{Membership}
\label{sec:membership}

We classified stars as members or non-members of their respective
clusters or galaxies on the basis of radial velocity and position in
the color--magnitude diagram.  For clusters, we additionally
considered the metallicity of the star.  K10 and K16 described
membership determination in detail for dwarf galaxies and GCs,
respectively.  In brief, the stars were selected to be within three
standard deviations of the mean radial velocity.  Stars were
eliminated if they were obviously non-members on the basis of their
colors or magnitudes or if they had strong Na$\,${\sc i} 8190
doublets, signaling that they are foreground dwarf stars.  Finally, GC
stars were selected to be within three standard deviations of the mean
metallicity.

Many stars were observed multiple times.  We use these repeat
observations in Section~\ref{sec:repeats} to validate our estimates of
uncertainty.  In all other contexts, we report the abundances for the
spectrum with the highest S/N\@.


\section{Elemental Abundance Measurements}
\label{sec:measure}

Traditional abundance measurements with equivalent widths are not
practical at the spectral resolution of DEIMOS\@.  Many lines are
blended together, making it unfeasible to disentangle the contribution
to line flux of one absorption line from another.  Therefore, we use
spectral synthesis.

K10 already measured atmospheric parameters and some elemental
abundances for the stars in their sample.  Because we are mostly using
the same spectra, we simply adopted their parameters.  We followed
exactly the same procedure as K10 to derive the parameters for the new
spectra described in Section~\ref{sec:newspec}.  The atmospheric
parameters are effective temperature ($T_{\rm eff}$), surface gravity
($\log g$), metallicity ([Fe/H]),\footnote{We use the notation ${\rm
    [A/B]} = \log[n({\rm A})/n({\rm B})] - \log[n_{\sun}({\rm
      A})/n_{\sun}({\rm B})]$, where $n({\rm A})$ is the number
  density of atom A\@.} and the $\alpha$-to-iron ratio
([$\alpha$/Fe]).  The last parameter encapsulates the amount of
various $\alpha$ elements, like Mg, that are present in the
atmosphere.  This parameter has a secondary effect on the atmospheric
structure of the star, largely through the presence of free electrons
donated by $\alpha$ elements, especially Mg \citep{van12}.  In
addition to Fe, individual elemental abundances of Mg, Si, Ca, and Ti
were also measured.

These atmospheric parameters and elemental abundances were measured
via $\chi^2$ minimization between the observed spectra and a large
grid of synthetic spectra.  The grid was synthesized at
0.02~\AA\ resolution over the spectral range 6300--9100~\AA\ at a
spacing of $\Delta T_{\rm eff} = 100$~K, $\Delta \log g = 0.5$,
$\Delta {\rm [Fe/H]} = 0.1$, and $\Delta {\rm [\alpha/Fe]} = 0.1$.
The spectra were continuum-normalized iteratively.

K10 used the atomic and molecular line list compiled by
\citet{kir08a}.  We used a subset of the same line list for the
measurement of iron-peak abundances.  Section~\ref{sec:linelist}
further describes the line list and our procedure for determining
which lines to use.

We adopted the solar abundances of \citet{asp09} except in the case of
Fe and Mg.  We used $\epsilon({\rm Fe}) = 12 + \log[n({\rm Fe})/n({\rm
    H})] = 7.52$ and $\epsilon({\rm Mg}) = 7.58$ for consistency with
K10.  The adopted iron-peak solar abundances are $\epsilon({\rm Cr}) =
5.64$, $\epsilon({\rm Co}) = 4.99$, and $\epsilon({\rm Ni}) = 6.22$.

For each star, we interpolated a synthetic spectrum from K10's grid
based on the previously determined atmospheric parameters.  This
spectrum has the solar ratios of [Cr/Fe], [Co/Fe], and [Ni/Fe].  We
call this spectrum $f_{\rm baseline}(\lambda_i)$, where $i$ is the
index of one pixel.  The spectrum was smoothed to match the resolution
of the DEIMOS spectrum.  This spectrum was used to determine the
continuum in the same manner as \citet{kir09} and K10, by fitting a
spline to the quotient of the observed spectrum and its corresponding
model spectrum.  The division removes the absorption lines and gives a
mostly featureless spectrum, i.e., the continuum.  We divided the
observed spectrum by the continuum, yielding a spectrum normalized to
unity between absorption lines.

In order to measure the abundances of iron-peak elements, we isolated
the spectral regions sensitive to changes in the abundances of these
elements, as described in Section~\ref{sec:linelist}.  We synthesized
spectra in these small regions, including a 2~\AA\ buffer on both
sides of each region.  We used the 2017 version of
MOOG\footnote{Downloaded from
  \url{https://github.com/alexji/moog17scat}.} \citep{sne73}.  MOOG is
a spectral synthesis code that solves the radiative transfer equations
in local thermodynamic equilibrium (LTE)\@.  We used the version of
the code that was updated to include electron scattering in the
calculation of the continuum opacity \citep{sob11}, coupled with plane
parallel ATLAS9 \citep{kur93} model atmospheres computed in LTE by
\citet{kir11d}.  These spectra are called $f_{\rm mod}(\lambda_j)$,
where $j$ represents the indices of pixels in the iron-peak regions.
We then constructed new spectra, $f_{\rm merged}(\lambda_i)$, where
$f_{\rm merged}(\lambda_i) = f_{\rm mod}(\lambda_j)$ if $i \in j$ and
$f_{\rm merged}(\lambda_i) = f_{\rm baseline}(\lambda_i)$ otherwise.

\begin{splitdeluxetable*}{llccCCCClRCCBCCCCCCCC}
\tablewidth{0pt}
\tablecolumns{20}
\tablecaption{Catalog of Abundances\label{tab:catalog}}
\tablehead{\colhead{System} & \colhead{Star Name} & \colhead{R.A.\ (J2000)} & \colhead{Decl.\ (J2000)} & \colhead{$B$} & \colhead{$V$} & \colhead{$R$} & \colhead{$I$} & \colhead{Slitmask} & \colhead{S/N} & \colhead{$T_{\rm eff}$} & \colhead{$\log g$} & \colhead{[Fe/H]} & \colhead{[Mg/Fe]} & \colhead{[Si/Fe]} & \colhead{[Ca/Fe]} & \colhead{[Ti/Fe]} & \colhead{[Cr/Fe]} & \colhead{[Co/Fe]} & \colhead{[Ni/Fe]} \\
\colhead{ } & \colhead{ } & \colhead{ } & \colhead{ } & \colhead{(mag)} & \colhead{(mag)} & \colhead{(mag)} & \colhead{(mag)} & \colhead{ } & \colhead{(\AA$^{-1}$)} & \colhead{(K)} & \colhead{(cm~s$^{-2}$)} & \colhead{ } & \colhead{ } & \colhead{ } & \colhead{ } & \colhead{ } & \colhead{ } & \colhead{ } & \colhead{ }}
\startdata
NGC 288           & N288-S56           & 00 52 35.93 & -26 37 40.5 &   18.17 &   17.52 & \nodata &   16.72 & n288    &  54.3 & 5400 & 3.35 & -1.24 \pm 0.11 & +0.24 \pm 0.27 & +0.27 \pm 0.18 & +0.60 \pm 0.18 & +0.18 \pm 0.23 &    \nodata     & +0.52 \pm 0.27 & -0.02 \pm 0.16 \\
NGC 288           & 10795              & 00 52 36.08 & -26 37 06.3 & \nodata &   14.79 & \nodata &   13.73 & 288l3   & 261.4 & 4738 & 1.93 & -1.42 \pm 0.10 & +0.41 \pm 0.09 & +0.42 \pm 0.10 & +0.12 \pm 0.12 & +0.30 \pm 0.11 & -0.12 \pm 0.14 & +0.04 \pm 0.17 & -0.06 \pm 0.08 \\
NGC 288           & N288-S57           & 00 52 36.29 & -26 34 07.8 &   17.99 &   17.33 & \nodata &   16.52 & n288    &  59.1 & 5351 & 3.25 & -1.14 \pm 0.11 & +0.48 \pm 0.21 & +0.21 \pm 0.16 & +0.02 \pm 0.21 & +0.24 \pm 0.17 &    \nodata     & +0.39 \pm 0.26 & -0.29 \pm 0.14 \\
NGC 288           & 10740              & 00 52 36.66 & -26 37 23.8 & \nodata &   16.82 & \nodata &   15.98 & 288l1   & 110.2 & 5118 & 3.01 & -1.43 \pm 0.10 & +0.30 \pm 0.15 & +0.11 \pm 0.18 & +0.17 \pm 0.17 & +0.06 \pm 0.14 & -0.11 \pm 0.14 & +0.11 \pm 0.30 & -0.21 \pm 0.13 \\
NGC 288           & 905                & 00 52 36.74 & -26 38 29.2 & \nodata &   17.19 & \nodata &   16.38 & 288l2   & 136.1 & 5327 & 3.19 & -1.29 \pm 0.10 & +0.33 \pm 0.12 & +0.46 \pm 0.11 & +0.25 \pm 0.12 & +0.31 \pm 0.12 & -0.32 \pm 0.15 & +0.20 \pm 0.19 & -0.05 \pm 0.09 \\
NGC 288           & N288-S155          & 00 52 36.86 & -26 30 58.6 &   15.32 &   14.37 & \nodata &   13.33 & n288    & 265.2 & 4640 & 1.79 & -1.32 \pm 0.10 & +0.23 \pm 0.09 & +0.49 \pm 0.10 & +0.27 \pm 0.12 & +0.31 \pm 0.11 & -0.14 \pm 0.13 & +0.09 \pm 0.17 & +0.00 \pm 0.08 \\
NGC 288           & 10554              & 00 52 37.14 & -26 38 57.3 & \nodata &   17.51 & \nodata &   16.69 & 288l5   & 125.2 & 5266 & 3.32 & -1.31 \pm 0.10 & +0.33 \pm 0.11 & +0.28 \pm 0.11 & +0.19 \pm 0.12 & +0.18 \pm 0.13 & -0.26 \pm 0.28 & -0.11 \pm 0.26 & -0.16 \pm 0.09 \\
NGC 288           & 10922              & 00 52 37.16 & -26 36 26.1 & \nodata &   16.19 & \nodata &   15.27 & 288l2   & 193.6 & 5079 & 2.68 & -1.36 \pm 0.10 &    \nodata     & +0.42 \pm 0.10 & +0.28 \pm 0.12 & +0.23 \pm 0.11 & -0.07 \pm 0.15 & +0.14 \pm 0.18 & -0.08 \pm 0.09 \\
NGC 288           & N288-S53           & 00 52 37.34 & -26 38 11.5 &   18.12 &   17.47 & \nodata &   16.66 & n288    &  56.6 & 5390 & 3.32 & -1.16 \pm 0.11 & +0.74 \pm 0.15 & +0.20 \pm 0.17 & +0.19 \pm 0.19 & +0.28 \pm 0.20 &    \nodata     & +0.69 \pm 0.24 & -0.19 \pm 0.16 \\
NGC 288           & 385                & 00 52 37.44 & -26 39 28.3 & \nodata &   16.48 & \nodata &   15.61 & 288l1   & 122.7 & 5160 & 2.84 & -1.24 \pm 0.10 & +0.43 \pm 0.14 & +0.43 \pm 0.11 & +0.18 \pm 0.12 & +0.22 \pm 0.12 & +0.06 \pm 0.17 & +0.25 \pm 0.20 & -0.33 \pm 0.09 \\
\enddata
\tablecomments{Table~\ref{tab:catalog} is published in its entirety in the machine-readable format.  A portion is shown here for guidance regarding its form and content.}
\end{splitdeluxetable*}

We measured elemental abundances by varying the input abundance in the
computation of the spectra and calculating $\chi^2$ between the
observed and synthetic spectra.  This differs slightly from the
approach of K10 in that we synthesized small regions of the spectrum
on the fly rather than pre-computing a grid of synthetic spectra.  We
found the minimum $\chi^2$ with Levenberg--Marquardt minimization
accomplished via the IDL routine MPFIT \citep{mar12}.  We quote the
abundance corresponding to the minimum $\chi^2$ as the measured
abundance.  Table~\ref{tab:catalog} gives the coordinates, photometric
magnitudes, slitmask, S/N, temperature, gravity, and Fe, Mg, Si, Ca,
Ti, Cr, Co, and Ni abundances for each star in the sample.\footnote{In
  the course of preparing this article, we found a minor bug in the
  code that determines the spectral resolution of the DEIMOS spectrum.
  The abundances reported here were computed after fixing this bug.
  The typical change for any of the reported abundances is $\pm
  0.03$~dex.}  Abundances with uncertainty estimates greater than
0.3~dex (as determined in Section~\ref{sec:errors}) are not shown.

\subsection{Iron-peak Line List}
\label{sec:linelist}

\begin{deluxetable}{llcc}
\tablewidth{0pt}
\tablecolumns{4}
\tablecaption{Important Atomic Lines\label{tab:linelist}}
\tablehead{\colhead{Wavelength} & \colhead{Species} & \colhead{Excitation Potential} & \colhead{$\log gf$} \\
\colhead{(\AA, air)} & \colhead{ } & \colhead{(eV)} & \colhead{ }}
\startdata
7290.864 & \ion{Ni}{1} & 5.342 &     $-0.235$ \\
7291.453 & \ion{Ni}{1} & 1.935 &     $-2.700$ \\
7355.890 & \ion{Cr}{1} & 2.890 &     $-0.100$ \\
7393.600 & \ion{Ni}{1} & 3.606 &     $-0.150$ \\
7400.180 & \ion{Cr}{1} & 2.900 & \phs$ 0.050$ \\
7409.346 & \ion{Ni}{1} & 3.796 &     $-0.100$ \\
7414.500 & \ion{Ni}{1} & 1.986 &     $-2.300$ \\
7417.267 & \ion{Co}{1} & 2.042 &     $-2.900$ \\
7417.330 & \ion{Co}{1} & 2.042 &     $-2.500$ \\
7417.350 & \ion{Co}{1} & 2.042 &     $-3.100$ \\
\enddata
\tablecomments{Table~\ref{tab:linelist} is published in its entirety in the machine-readable format.  A portion is shown here for guidance regarding its form and content.}
\end{deluxetable}

\begin{figure*}
\includegraphics[width=0.27623\linewidth]{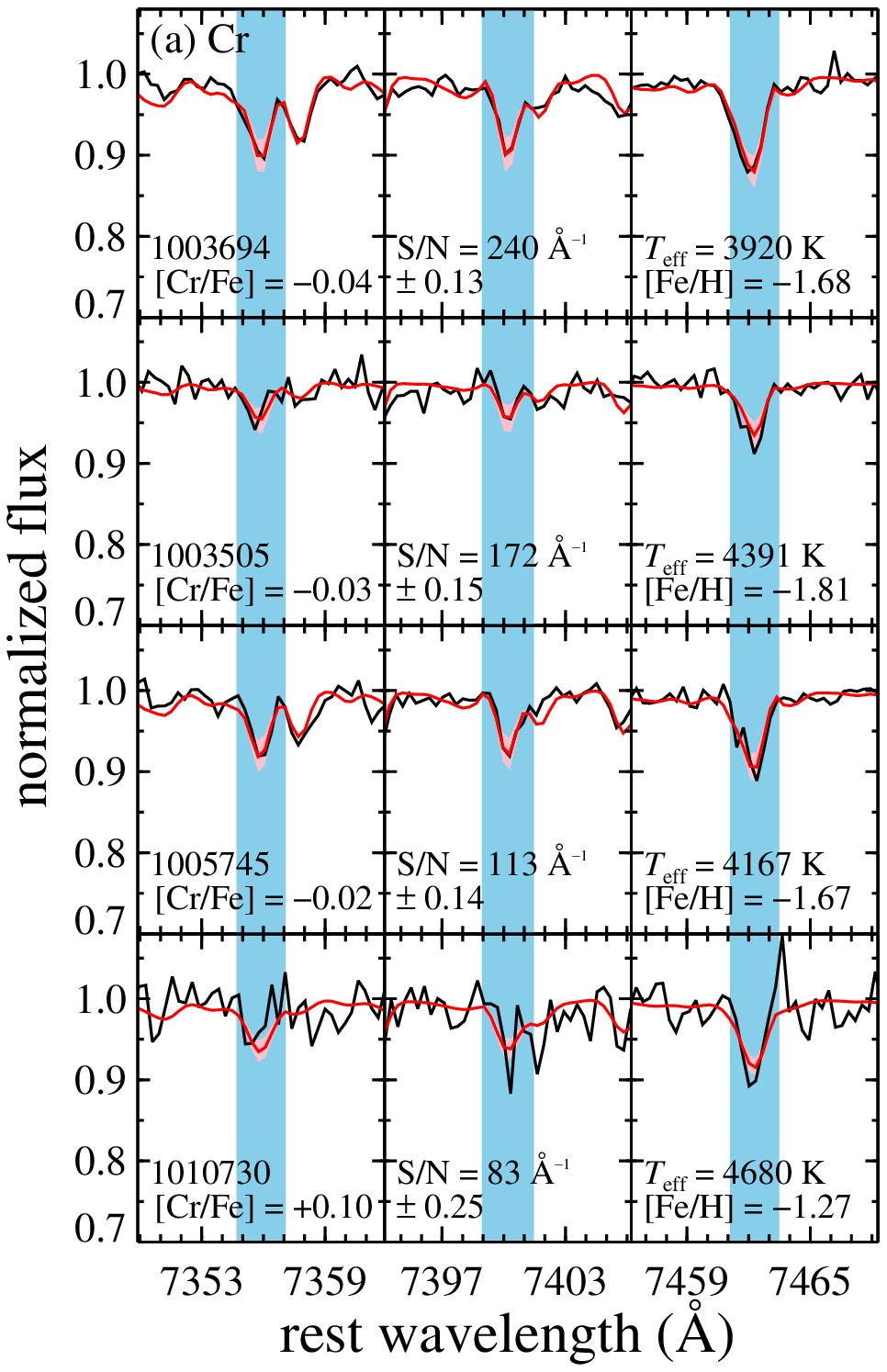}
\hfill
\includegraphics[width=0.35189\linewidth]{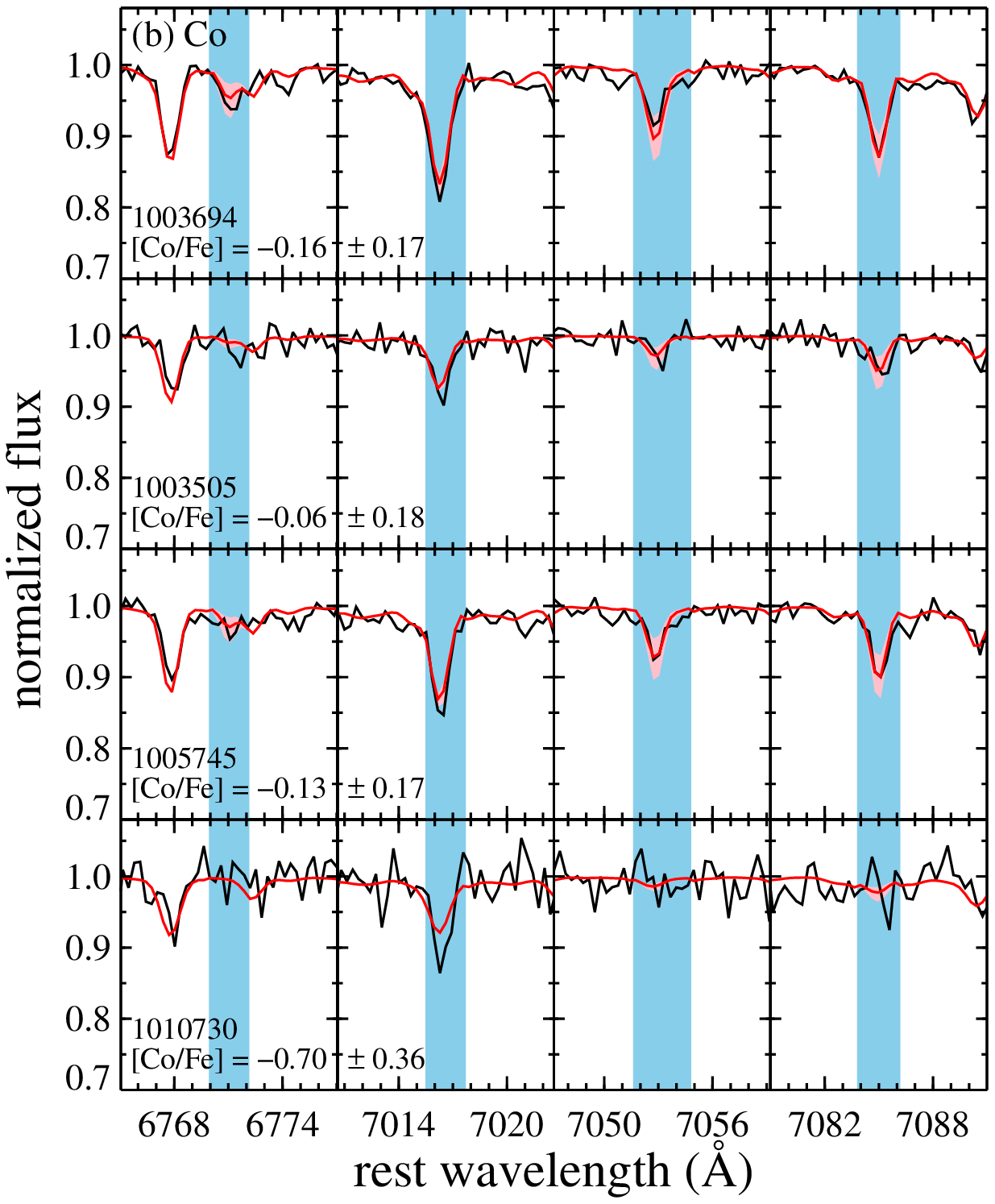}
\hfill
\includegraphics[width=0.35189\linewidth]{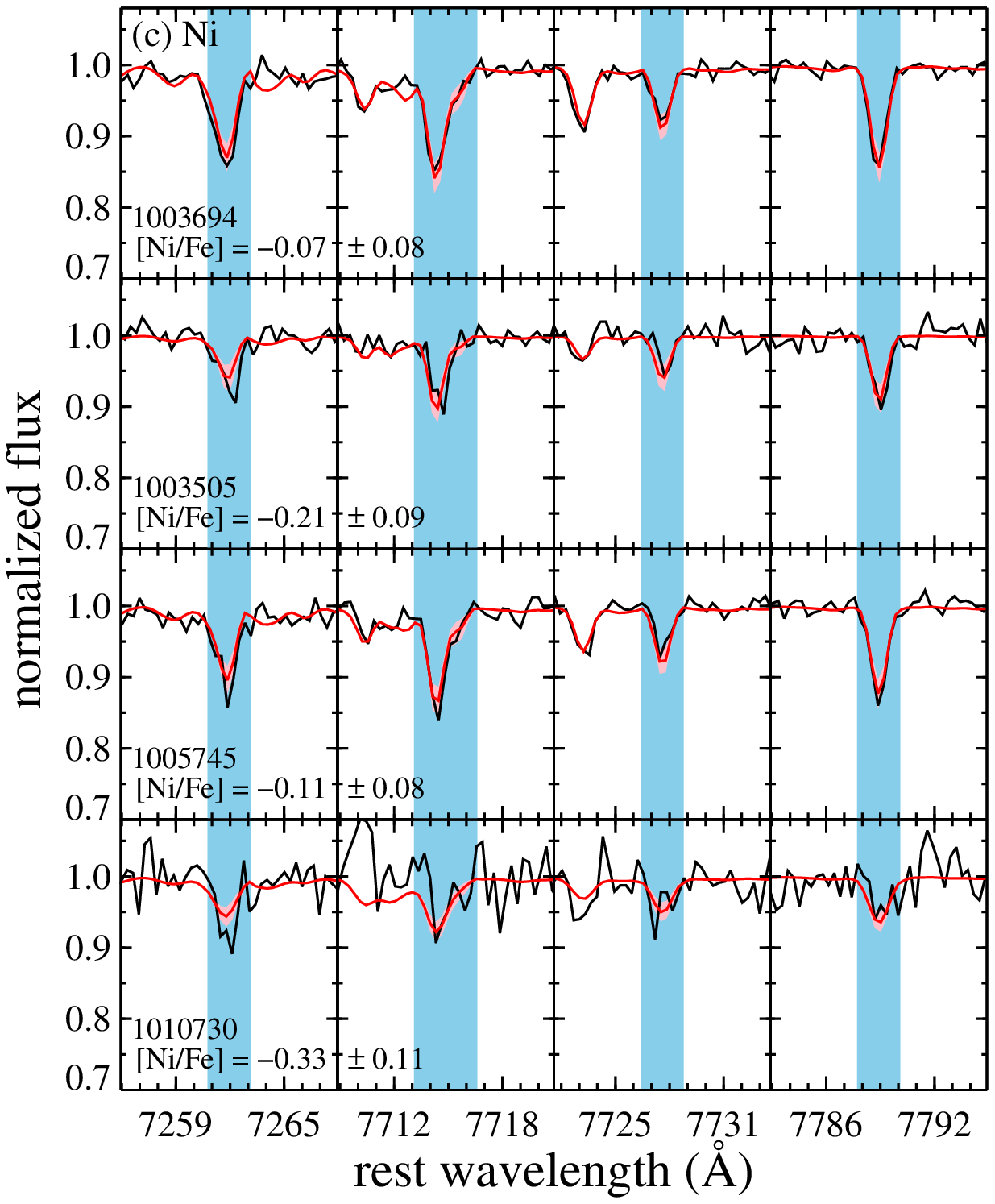}
\caption{Small regions of the DEIMOS spectra showing absorption
  features of (a) Cr, (b) Co, and (c) Ni.  The blue shaded regions
  show spectral regions used in the calculation of $\chi^2$ (see
  Section~\ref{sec:linelist}).  The black curves show the observed
  spectra, and the red curves show the best-fitting model spectra.
  The pink shaded regions show the changes in the model spectra for
  changes of $\pm 0.3$~dex in abundance.  The same four stars are
  shown in each panel.  They are all members of Sculptor, and they are
  chosen to have high S/N (top) and moderately low S/N (bottom).
  Panel (a) gives $T_{\rm eff}$, [Fe/H], and S/N for each star.  The
  [Co/Fe] measurement in the bottom panel does not pass our error
  threshold of 0.3~dex.  As a result, the Co measurement for this star
  does not appear in the catalog (Table~\ref{tab:catalog}), but we
  plot the spectrum here anyway.\label{fig:spec}}
\end{figure*}

The line list used for the computation of all model spectra in this
work was taken from \citet{kir08a}, who compiled atomic and molecular
information from the Vienna Atomic Line Database
\citep[VALD,][]{kup99}, supplemented by molecular transitions
\citep{kur92}.  Additionally, single-line atomic data for select
elements, including Co, were replaced by hyperfine blends
\citep{kur93hyperfine}.  \citet{kir08a} modified the oscillator
strengths for some lines to match the spectra of the Sun and Arcturus.

Iron-peak elements other than iron do not affect the great majority of
the pixels in the DEIMOS spectra.  Therefore, we synthesized only
those small regions sensitive to iron-peak abundances in order to save
computation time.  We chose these regions by constructing model
spectra between $T_{\rm eff} = 3500$~K and 8000~K at a spacing of 50~K
and between $\log g = 1.5$ and 3.5 at a spacing of 0.5.  The
metallicity was ${\rm [Fe/H]} = -1.0$ for all of the models, and the
microturbulence was $\xi = 2.13 - 0.23 (\log g)$ \citep{kir09}.  The
model spectra were generated with MOOG and ATLAS9 model atmospheres,
just as before.  They were smoothed to match the DEIMOS spectral
resolution (line profiles with $\sim 1.2$~\AA\ FWHM)\@.  We separately
analyzed the effect of the iron-group elements Sc, V, Cr, Mn, Co, Ni,
Cu, and Zn on the model spectra.  (K10 already measured Ti.)  For each
element and each pair of $T_{\rm eff}$ and $\log g$, we synthesized
one spectrum with the solar ratio of the iron-peak element (e.g.,
${\rm [Ni/Fe]} = 0$) and one spectrum without any of the iron-peak
element (e.g., ${\rm [Ni/Fe]} = -\infty$).  We identified all the
spectral regions of at least 0.75~\AA\ in width where the flux changed
by at least 0.5\% for any of the pairs of $T_{\rm eff}$ and $\log g$.
Finally, we examined each spectral region to compare model spectra to
observed spectra with high S/N to verify that the model was a
reasonable representation of real spectra.  We discarded a few
spectral regions where the line strengths were obviously inconsistent
between the model and observed spectra.  The remaining spectral
windows range in size from 1.1 to 4.5~\AA\@.

Most elements did not have a large enough presence in the spectral
region under consideration (approximately 6300--9100~\AA) to be useful
for measuring their abundances.  Only Cr, Co, and Ni had enough
spectral regions with sufficiently large responses to changes in
abundance.  The remaining discussion of iron-peak elements will be
limited to these three elements.  Sc, V, Mn, Cu, and Zn are excluded
from the discussion.

Table~\ref{tab:linelist} gives the Cr, Co, and Ni absorption lines in
the spectral regions, including hyperfine structure for Co.  There are
9 Cr lines, 12 Co lines (140 when accounting for hyperfine structure),
and 34 Ni lines.  The $\chi^2$ measurement for each element was
performed only in the spectral regions sensitive to that element.  The
model spectra used for the computation of $\chi^2$ were computed with
the full line list, not just the lines in Table~\ref{tab:linelist}.

Figure~\ref{fig:spec} shows four examples of spectral regions for each
of Cr, Co, and Ni.  The number of spectral regions used exceeds those
shown in the figure.  The pink shading shows the sensitivity to
abundance changes of $\pm 0.3$~dex.  Although it may not seem that the
spectra are particularly sensitive to $\pm 0.3$~dex changes, the full
spectral range including all of the spectral regions has enough
sensitivity to justify the uncertainties we quote.

\subsection{Non-LTE Corrections}
\label{sec:nlte}

The assumption of LTE is sufficient for many applications.  For
example, iron abundances computed in LTE with one-dimensional
atmospheres are typically different from non-LTE (NLTE) computations
by $\lesssim 0.1$~dex \citep{ber12}.  However, electronic transitions
of some elements have larger NLTE corrections.  \citet{berces10} and
\citet{ber10} found that the NLTE corrections for Cr and Co are
significant.  Therefore, we explored the effect of including NLTE
corrections on our measurements of [Cr/Fe] and [Co/Fe].

We computed corrections based on the formalism of \citet{berces10} and
\citet{ber10}, which we briefly describe here.  The corrections were
computed with two sets of model atmospheres: one-dimensional, plane
parallel models generated with the MAFAGS-OS code
\citep{gru04a,gru04b} and one-dimensional, spherical models generated
with the MARCS code \citep{gus08}.  We used the NLTE corrections based
on the MARCS grid because they were computed in a temperature range
that overlaps our sample better than the MAFAGS-OS grid.  The
corrections were computed on a grid of atmospheric parameters spanning
$2500~{\rm K} \le T_{\rm eff} \le 7750~{\rm K}$, $-0.5 \le \log g \le
3.5$, and $-5.0 \le {\rm [Fe/H]} \le 1.0$.  Microturbulence was fixed
at $\xi = 2.13 - 0.23 (\log g)$ \citep{kir09}.  An NLTE spectrum with
${\rm [Cr/Fe]} = 0$ or ${\rm [Co/Fe]} = 0$ was computed for each point
on this grid for all Cr and Co transitions in
Table~\ref{tab:linelist}.  The syntheses were 10~\AA\ wide centered on
each line.  Additionally, 21 LTE spectra were computed with abundance
corrections ranging from $\Delta \epsilon = -1.0$ to $1.0$ with a step
size of $0.1$.  Equivalent widths were computed for all spectra.  The
NLTE correction was taken to be the change in LTE abundance required
to make the LTE equivalent width match the NLTE equivalent width.  We
performed linear interpolation in $\Delta \epsilon$ to minimize the
difference in equivalent width between the LTE and NLTE spectra.
These corrections are available in a queryable online
database.\footnote{\url{http://nlte.mpia.de}\label{note:nlte}}

Because the NLTE corrections are different for each transition, we
cannot apply a single NLTE correction to the Cr or Co abundance of
each star.  Instead, we must apply separate NLTE corrections to the
synthesis of each transition.  We followed the same procedure
described in Section~\ref{sec:measure} except that we altered the
oscillator strength ($\log gf$) of each Cr or Co transition.  We
changed the oscillator strength by the abundance correction described
in the previous paragraph.  The effect of changing oscillator strength
is identical to changing the input abundance.  However, our approach
allows us to compute synthetic spectra with multiple Cr or Co
transitions, each with separate NLTE corrections.  The new oscillator
strength is $\log gf_{{\rm mod},i} = \log gf_{{\rm orig},i} - \Delta
\epsilon_i$.  In this equation, $\log gf_{{\rm mod},i}$ is the
oscillator strength for line $i$ to be used as the input to an LTE
code, like MOOG\@.  Finally, $\Delta \epsilon_i$ is the NLTE
correction for line $i$, which is a function of temperature, gravity,
and metallicity.  The sign convention is such that $\Delta \epsilon_i$
is the value that should be added to an LTE Co abundance measurement
to give the NLTE abundance.  This is the same convention on the NLTE
website in footnote~\ref{note:nlte}.

The resulting spectra had 9 Cr and 12 Co transitions.  These were the
model spectra that were used in the $\chi^2$ minimization.  This
procedure required us to sample NLTE corrections both between grid
points and beyond the bounds of the NLTE corrections grid.  We used
linear interpolation and, where necessary, extrapolation.

\begin{figure}
\includegraphics[width=\linewidth]{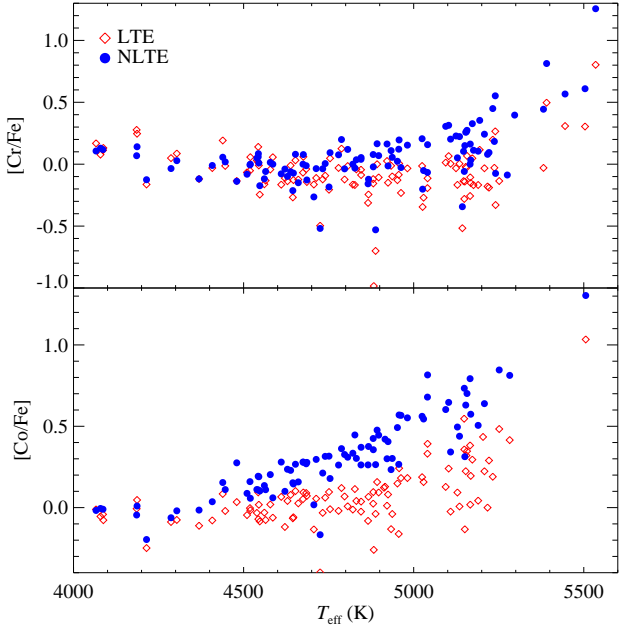}
\caption{LTE and NLTE abundance ratios for [Cr/Fe] and [Co/Fe] as
  function of effective temperature in the globular cluster
  M79.\label{fig:nlte}}
\end{figure}

Figure~\ref{fig:nlte} shows the LTE and NLTE abundances for the
globular cluster M79 as a function of $T_{\rm eff}$.  We expect that
the Cr, Fe, and Co abundances in M79 are constant.  Therefore, they
should not have a trend with $T_{\rm eff}$.  Any trend would indicate
a systematic error in the abundances, possibly caused by the
assumption of LTE\@.  The LTE abundances for Cr show a U shape with
$T_{\rm eff}$, whereas the LTE abundances for Co rise by about 0.1~dex
from $T_{\rm eff} = 4000$~K to 5000~K\@.  However, the NLTE abundances
increase strongly with $T_{\rm eff}$.  The increase for ${\rm
  [Cr/Fe]}_{\rm NLTE}$ is 0.5~dex from 4900~K to 5400~K, and the
increase in ${\rm [Co/Fe]}_{\rm NLTE}$ is 0.8~dex from 4200~K to
5200~K\@.

The strong trend of NLTE abundances with $T_{\rm eff}$ indicates that
our application of NLTE corrections does not lead to more accurate
abundances.  The cause for the lack of improvement is under
investigation.  One possibility is that the atmospheric parameters
($T_{\rm eff}$, $\log g$, [Fe/H], and $\xi$) were computed assuming
LTE\@.  It is possible that the NLTE corrections are not appropriate
for atmospheric parameters derived assuming LTE\@.

In summary, our investigation into NLTE corrections found that the
corrections do not improve the abundance measurements.  Therefore, we
report only LTE abundances.  The magnitude of the deviation of the LTE
measurements of [Cr/Fe] and [Co/Fe] with $T_{\rm eff}$
(Figure~\ref{fig:nlte}) provides one estimate for the magnitude of
errors introduced by assuming LTE\@.  In Section~\ref{sec:errors}, we
estimate the the bulk effect of systematic errors by quantifying the
scatter of abundances within a globular cluster.  To the extent that
assuming LTE increases this scatter, as shown in
Figure~\ref{fig:nlte}, the error introduced by assuming LTE will be
included in those estimates.


\section{Estimation of Uncertainties}
\label{sec:errors}

The uncertainty on the abundance ratios, $\delta{\rm [X/Fe]}$,
consists of two components: a random uncertainty from the spectral
fitting and a systematic error.  Our error model assumes that these
components add in quadrature.

\begin{equation}
  \delta{\rm [X/Fe]} = \sqrt{\delta_{\rm fit}{\rm [X/Fe]}^2 + \delta_{\rm sys}{\rm [X/Fe]}^2} \label{eq:error}
\end{equation}

\noindent
MPFIT calculates the random uncertainty, $\delta_{\rm fit}{\rm
  [X/Fe]}$, as the change in [X/Fe] required to increase the unreduced
$\chi^2$ by one.

We estimated the systematic error, $\delta_{\rm sys}{\rm [X/Fe]}$, in
a manner similar to that of \citet{kir08a}.  The method uses only
stars in GCs.  We assume that each GC has a single value of [Cr/Fe], a
single value of [Co/Fe], and a single value of [Ni/Fe].  Any deviation
in the abundance measurements of any one star from these values must
be explained by the quoted uncertainty in the star.  We accomplish
this by computing a value called $\Delta$:

\begin{equation}
  \Delta_i = \frac{{\rm [X/Fe]}_i - \langle {\rm [X/Fe]} \rangle_j}{\sqrt{\delta_{\rm fit}{\rm [X/Fe]}^2 + \delta_{\rm sys}{\rm [X/Fe]}^2}} \label{eq:deltasys}
\end{equation}

\noindent
where $i$ represents an individual star and $j$ represents the cluster
to which that star belongs.  The mean value for the $j^{\rm th}$
cluster, $\langle {\rm [X/Fe]} \rangle_j$, is weighted by the inverse
variances of the measurements of [X/Fe].

\begin{figure*}
\includegraphics[width=\linewidth]{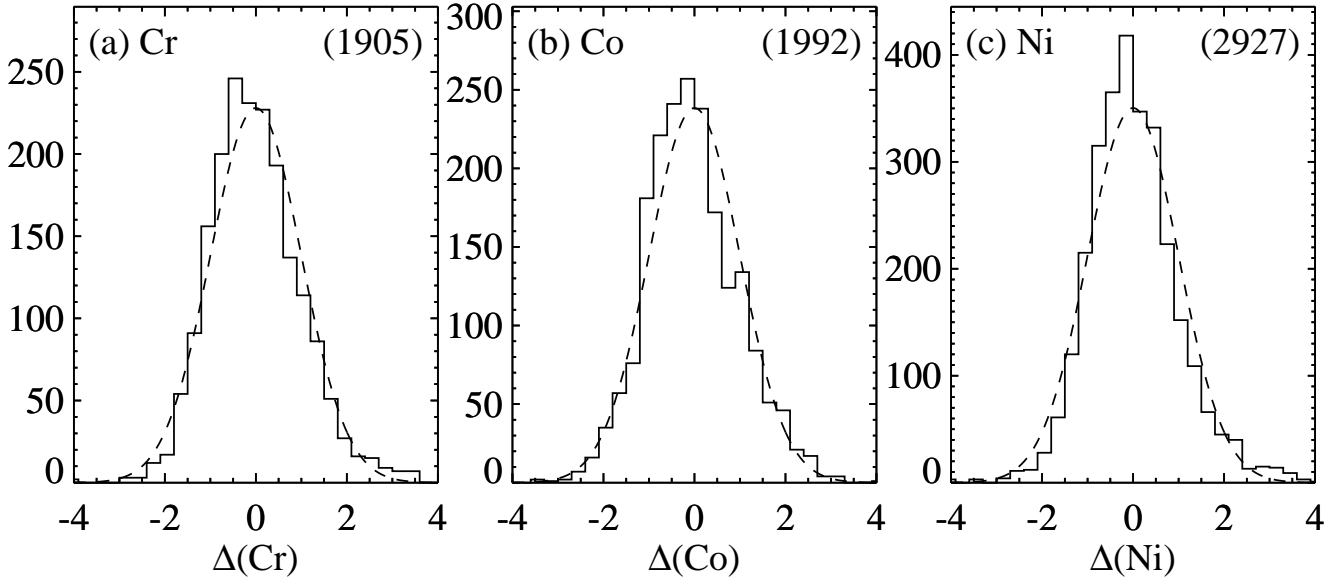}
\caption{Distributions of $\Delta$ (Equation~\ref{eq:deltasys}) used
  in determining the systematic errors for the iron-peak abundance
  measurements.  The dashed curves show Gaussians with a variance of
  unity.  In the limit of perfectly Gaussian-distributed
  uncertainties, the histograms would conform to the dashed curves.
  The numbers of stars are indicated in
  parentheses.\label{fig:syserr}}
\end{figure*}

We expect $\Delta$ to obey a Gaussian distribution with a variance of
unity if the uncertainties on [X/Fe] are properly estimated and
Gaussian-distributed.  Therefore, we determine $\delta_{\rm sys}{\rm
  [X/Fe]}$ by finding the value that gives $\Delta$ a variance of
unity.  The corresponding systematic errors for [Cr/Fe], [Co/Fe], and
[Ni/Fe], respectively, are \crsyserr, \cosyserr, and \nisyserr.

Figure~\ref{fig:syserr} shows the distribution of $\Delta$ for these
values of $\delta_{\rm sys}{\rm [X/Fe]}$.  The uncertainties shown in
Table~\ref{tab:catalog} are derived from Equation~\ref{eq:error}.  The
table includes only those stars with uncertainties less than 0.3~dex.
The median uncertainties are \medcrfeerr, \medcofeerr, and
\mednifeerr\ for [Cr/Fe], [Co/Fe], and [Ni/Fe], respectively.

Our estimate of the systematic error is predicated on the assumption
that the element ratios have no dispersion within globular clusters.
This assumption might be challenged by our inclusion of M2 and M22,
which are globular clusters that have been suggested to display
intrinsic dispersions in their iron abundances \citep{mar09,yon14}.
However, there is no evidence that either cluster shows a dispersion
in the ratios of an iron-peak element to iron, such as [Ni/Fe].
Furthermore, the claimed dispersion in iron has been challenged in
both cases as a side effect of using spectroscopic rather than
photometric surface gravities \citep{muc15a,lar16}.  In any case, we
experimented by determining $\delta_{\rm sys}{\rm [X/Fe]}$, excluding
M2 and M22.  The resulting values of $\delta_{\rm sys}{\rm [X/Fe]}$
actually increased by 0.003--0.008~dex, even though we would expect
them to decrease if the clusters had intrinsic dispersions in
iron-peak abundance ratios.  Therefore, we kept these clusters in the
sample.


\section{Validation}
\label{sec:validation}

We validate our measurements in two ways.  First, we quantify the
consistency among separate measurements of the same star.  In this
way, we assess the validity of the estimated uncertainties.  Second,
we compare our measurements to previously published high-resolution
spectroscopic abundance measurements of the same stars in order to
assess both accuracy and precision.

\subsection{Consistency Among Repeated Observations}
\label{sec:repeats}

One way to assess the resilience of our measurements to both spectral
noise and certain systematic errors is to examine repeat measurements
of the same star.  Our sample contains \ncrdup, \ncodup, and
\nnidup\ duplicate measurements of [Cr/Fe], [Co/Fe], and [Ni/Fe],
respectively.  These numbers include all the $\binom{N}{2}$
permutations for every star that was observed $N$ times.

We quantify the consistency among these repeat measurements in a
manner similar to Equation~\ref{eq:deltasys}.  For each pair of
observations $i$ and $j$, we compute the quantity $\delta_{ij}$:

\begin{equation}
  \delta_{ij} = \frac{{\rm [X/Fe]}_i - {\rm [X/Fe]}_j}{\sqrt{\delta_{\rm fit}{\rm [X/Fe]}_i^2 + \delta_{\rm fit}{\rm [X/Fe]}_j^2 + \delta_{\rm sys}{\rm [X/Fe]}^2}} \label{eq:deltadup}
\end{equation}

\noindent The denominator includes only one factor of the systematic
error term, even though $\delta_{ij}$ is computed from the
observations of two stars.  We made this choice because some of the
systematic error comes from variables, such as the photometry, that
are common to both measurements.  In a more sophisticated analysis,
the systematic error term would have a coefficient that encapsulates
the degree of dissimilarity between the variables that enter into the
computation of the abundance measurements for the pair of spectra.

\begin{figure*}
\includegraphics[width=\linewidth]{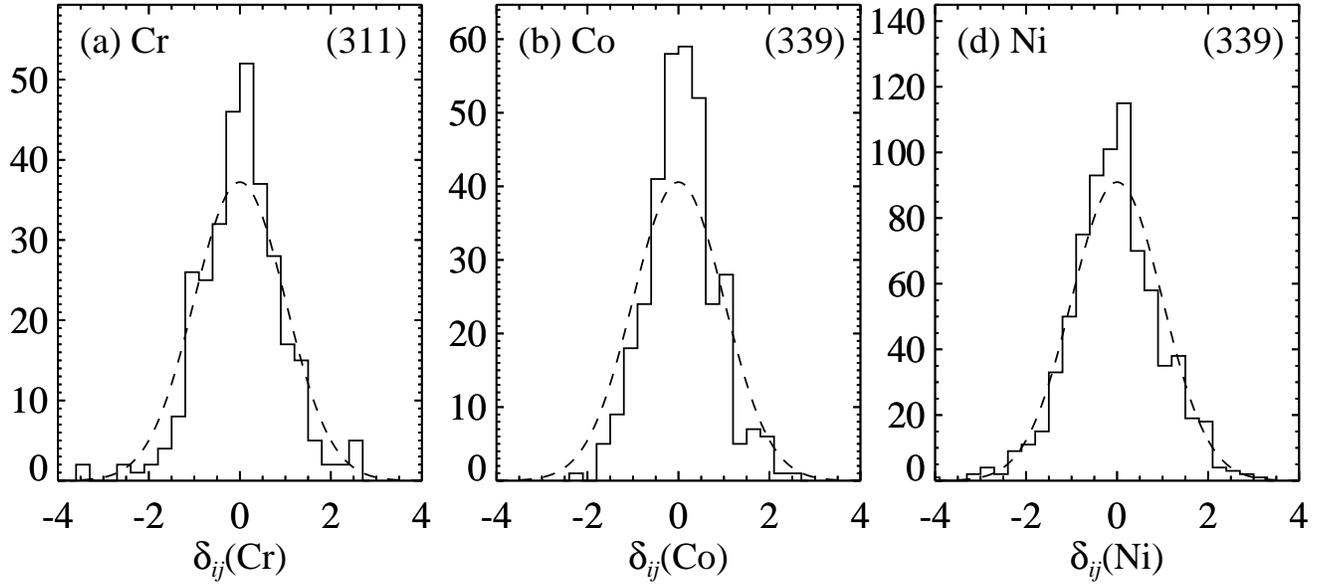}
\caption{Differences in measured abundance ratios between repeated
  observations of the same stars.  The differences are normalized by
  the quadrature sum of the uncertainties
  (Equation~\ref{eq:deltadup}).  The dashed curves show Gaussians with
  zero mean and unit variance.  Each panel gives the number of pairs
  of measurements in parentheses.
  \label{fig:repeats}}
\end{figure*}

Figure~\ref{fig:repeats} shows the distributions of $\delta_{ij}$ for
each element.  In the limit of perfectly determined abundances and
perfectly estimated uncertainties, the distribution of $\delta_{ij}$
would have a mean of zero and a variance of unity, as represented by
the dashed curves.  The actual mean values of $\delta_{ij}$ are
$\meancrdup \pm \meancrduperr$, $\meancodup \pm \meancoduperr$, and
$\meannidup \pm \meanniduperr$ for Cr, Co, and Ni, respectively.  In
reality, these mean values are not meaningful because the stars $i$
and $j$ in Equation~\ref{eq:deltadup} can be swapped arbitrarily.

The standard deviations of the $\delta_{ij}$ are $\sdcrdup \pm
\sdcrduperr$, $\sdcodup \pm \sdcoduperr$, and $\sdnidup \pm
\sdniduperr$ for Cr, Co, and Ni, respectively.  These values indicate
that we have estimated $\delta_{\rm sys}{\rm [Cr/Fe]}$ and
$\delta_{\rm sys}{\rm [Ni/Fe]}$ well, but we have perhaps
overestimated $\delta_{\rm sys}{\rm [Co/Fe]}$.  It could even be the
case that Co abundances are variable within GCs.  Our estimation of
uncertainties (Section~\ref{sec:errors}) assumes that each GC has a
single Co abundance.  If it is variable, then $\delta_{\rm sys}{\rm
  [Co/Fe]}$ will encompass that variability.  Alternatively, the
denominator in Equation~\ref{eq:deltadup} might not properly account
for the systematic error for duplicate measurements, as discussed
above.  For example, there seems to be a slight trend of abundance
with $T_{\rm eff}$ (Figure~\ref{fig:nlte}).  This systematic trend
would not be represented by $\delta_{ij}$ because the repeat
measurements have similar values of $T_{\rm eff}$.  In either case,
the errors do not appear to be underestimated.

\subsection{Comparison to High-resolution Spectroscopic Abundances}
\label{sec:validate}

\begin{splitdeluxetable*}{lllBCCCCCC}
\tablewidth{0pt}
\tablecolumns{9}
\tablecaption{Comparison to High-resolution Spectroscopic Abundances\label{tab:compare}}
\tablehead{\colhead{System} & \colhead{Star Name} & \colhead{HRS Reference} & \colhead{${\rm [Cr/Fe]}_{\rm HRS}$} & \colhead{${\rm [Co/Fe]}_{\rm HRS}$} & \colhead{${\rm [Ni/Fe]}_{\rm HRS}$} & \colhead{${\rm [Cr/Fe]}_{\rm MRS}$} & \colhead{${\rm [Co/Fe]}_{\rm MRS}$} & \colhead{${\rm [Ni/Fe]}_{\rm MRS}$}}
\startdata
M68            & I-256        & \text{\citet{gra89}}       & +0.01 \pm 0.20 &    \nodata     & -0.16 \pm 0.20 & -0.10 \pm 0.14 &    \nodata     & -0.18 \pm 0.08 \\
M5             & III-149      & \text{\citet{iva01}}       &    \nodata     &    \nodata     & -0.06 \pm 0.03 &    \nodata     &    \nodata     & -0.07 \pm 0.08 \\
M5             & G18155\_0228 & \text{\citet{ram03}}       &    \nodata     &    \nodata     & -0.14 \pm 0.15 &    \nodata     &    \nodata     & -0.00 \pm 0.11 \\
M5             & 1-36         & \text{\citet{ram03}}       & +0.07 \pm 0.20 & +0.05 \pm 0.11 & -0.09 \pm 0.05 & -0.26 \pm 0.26 & +0.01 \pm 0.25 & -0.00 \pm 0.11 \\
M5             & I-71         & \text{\citet{iva01}}       &    \nodata     &    \nodata     & -0.08 \pm 0.08 &    \nodata     &    \nodata     & -0.08 \pm 0.10 \\
M5             & II-59        & \text{\citet{iva01}}       &    \nodata     &    \nodata     & -0.06 \pm 0.04 &    \nodata     &    \nodata     & +0.03 \pm 0.08 \\
M5             & I-58         & \text{\citet{iva01}}       &    \nodata     &    \nodata     & -0.02 \pm 0.11 &    \nodata     &    \nodata     & +0.03 \pm 0.08 \\
M5             & G18447\_0453 & \text{\citet{ram03}}       & -0.42 \pm 0.10 &    \nodata     & +0.06 \pm 0.19 & -0.30 \pm 0.15 &    \nodata     & -0.19 \pm 0.09 \\
M5             & 1-31         & \text{\citet{ram03}}       & -0.08 \pm 0.07 & -0.09 \pm 0.10 & -0.10 \pm 0.04 & -0.10 \pm 0.13 & -0.04 \pm 0.17 & -0.10 \pm 0.08 \\
M5             & IV-59        & \text{\citet{iva01}}       &    \nodata     &    \nodata     & -0.13 \pm 0.07 &    \nodata     &    \nodata     & +0.11 \pm 0.08 \\
\enddata
\tablecomments{Table~\ref{tab:compare} is published in its entirety in the machine-readable format.  A portion is shown here for guidance regarding its form and content.}
\end{splitdeluxetable*}

\nocite{coh05a}
\nocite{coh05b}
\nocite{coh09}
\nocite{coh10}
\nocite{fre10b}
\nocite{ful00}
\nocite{ful04}
\nocite{gei05}
\nocite{gra89}
\nocite{iva01}
\nocite{joh02}
\nocite{koc08}
\nocite{kra98}
\nocite{lee05}
\nocite{let10}
\nocite{mis03}
\nocite{ram02}
\nocite{ram03}
\nocite{sad04}
\nocite{she01}
\nocite{she03}
\nocite{she98}
\nocite{sne97}

\begin{figure*}
\includegraphics[width=0.32666\linewidth]{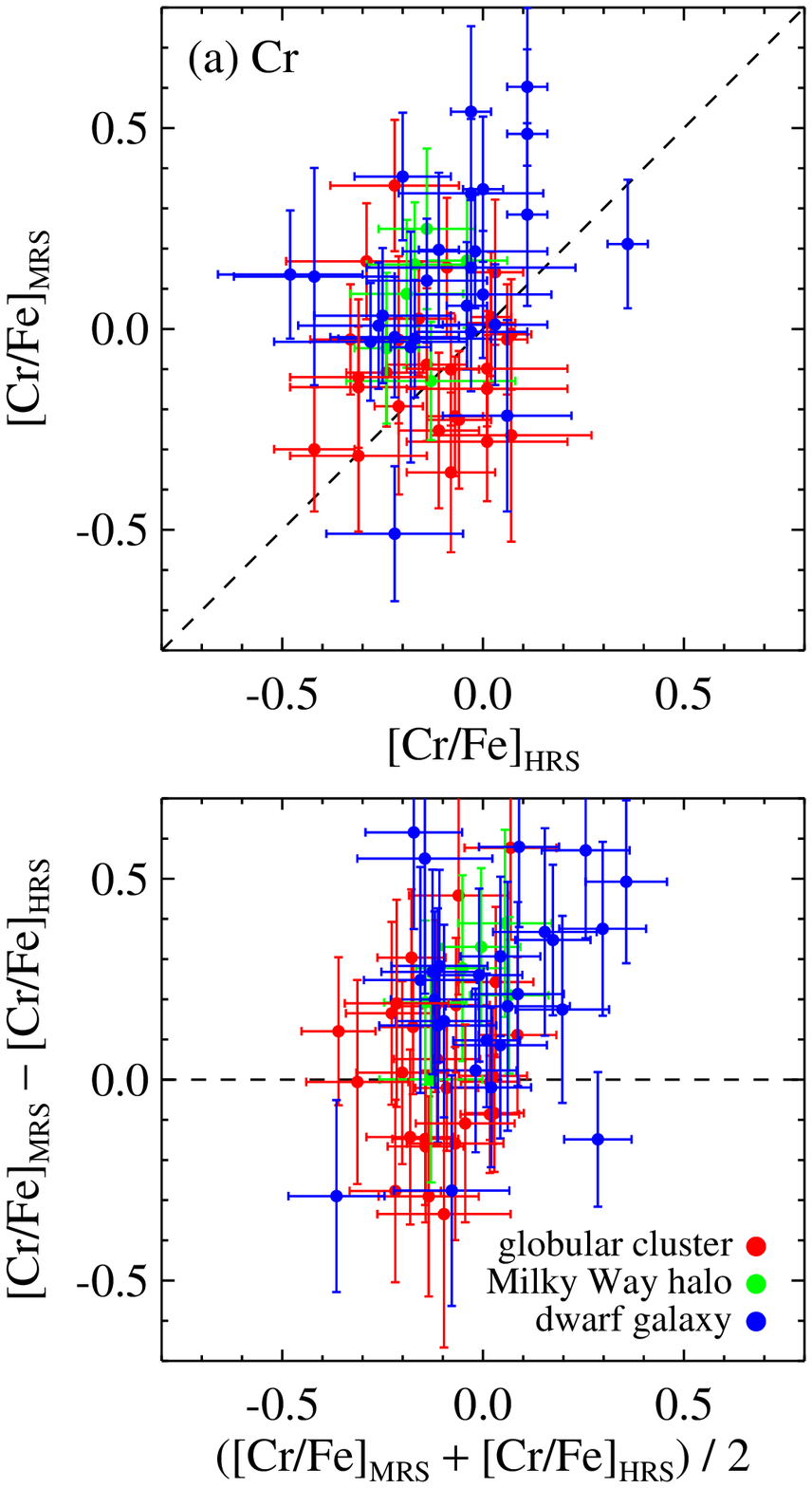}
\hfill
\includegraphics[width=0.32666\linewidth]{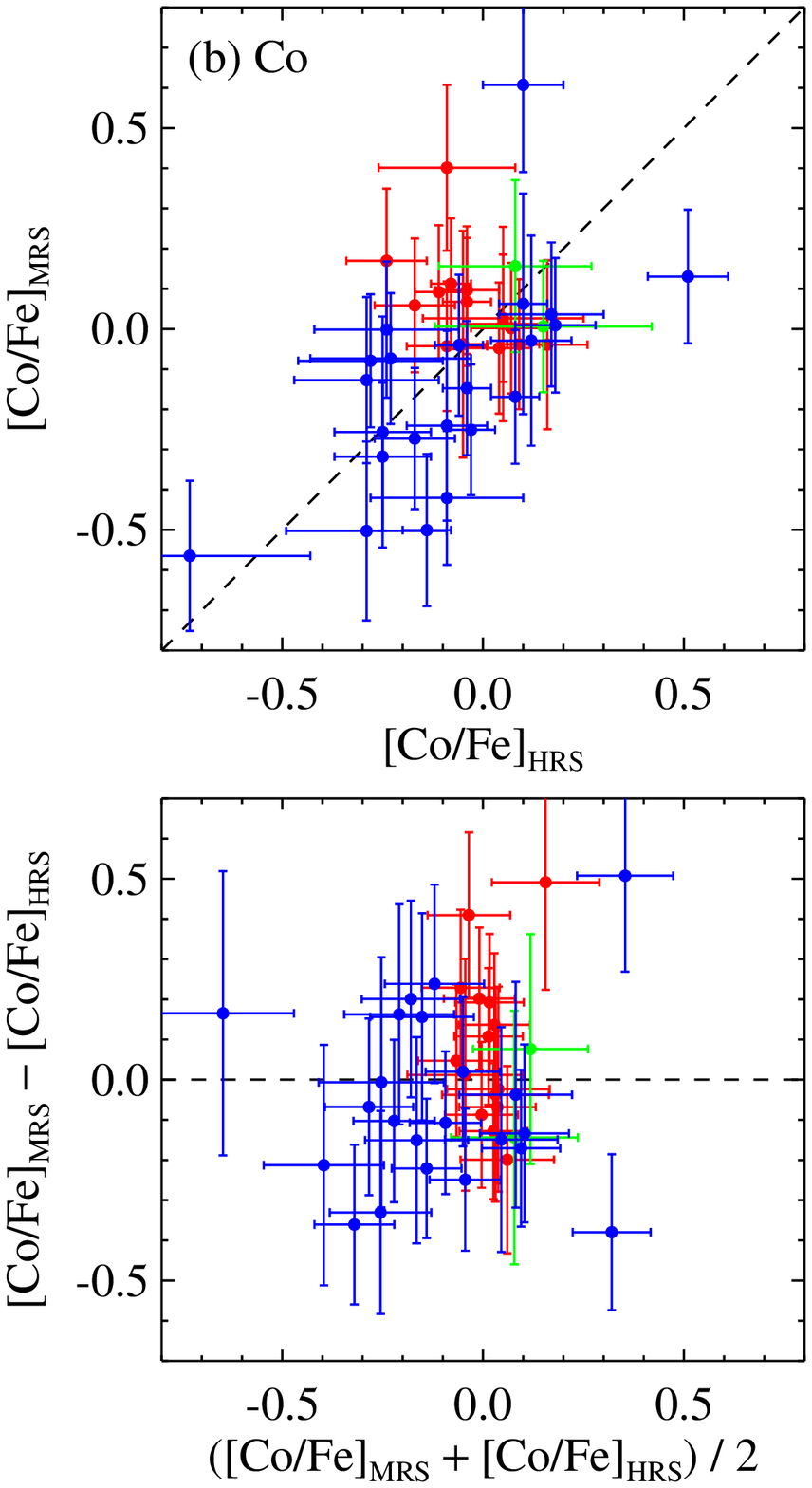}
\hfill
\includegraphics[width=0.32666\linewidth]{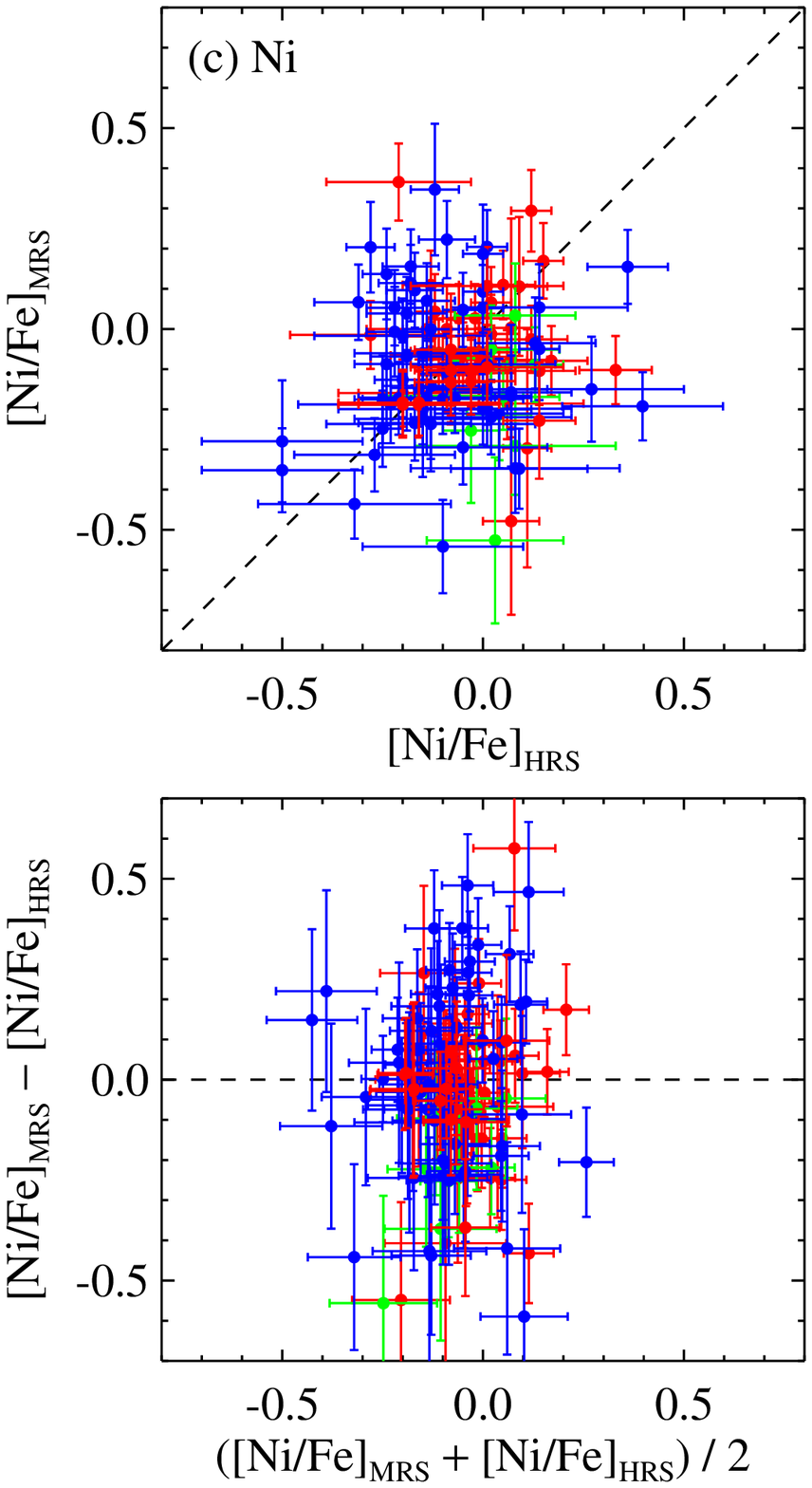}
\caption{Comparison between our MRS measurements and HRS measurements
  from the literature for the same stars for (a) [Cr/Fe], (b) [Co/Fe],
  and (c) [Ni/Fe].  Color-coding indicates the type of stellar system
  in which the stars reside.  The dashed lines indicate one-to-one
  agreement.\label{fig:validate}}
\end{figure*}

\begin{figure*}
\includegraphics[width=\linewidth]{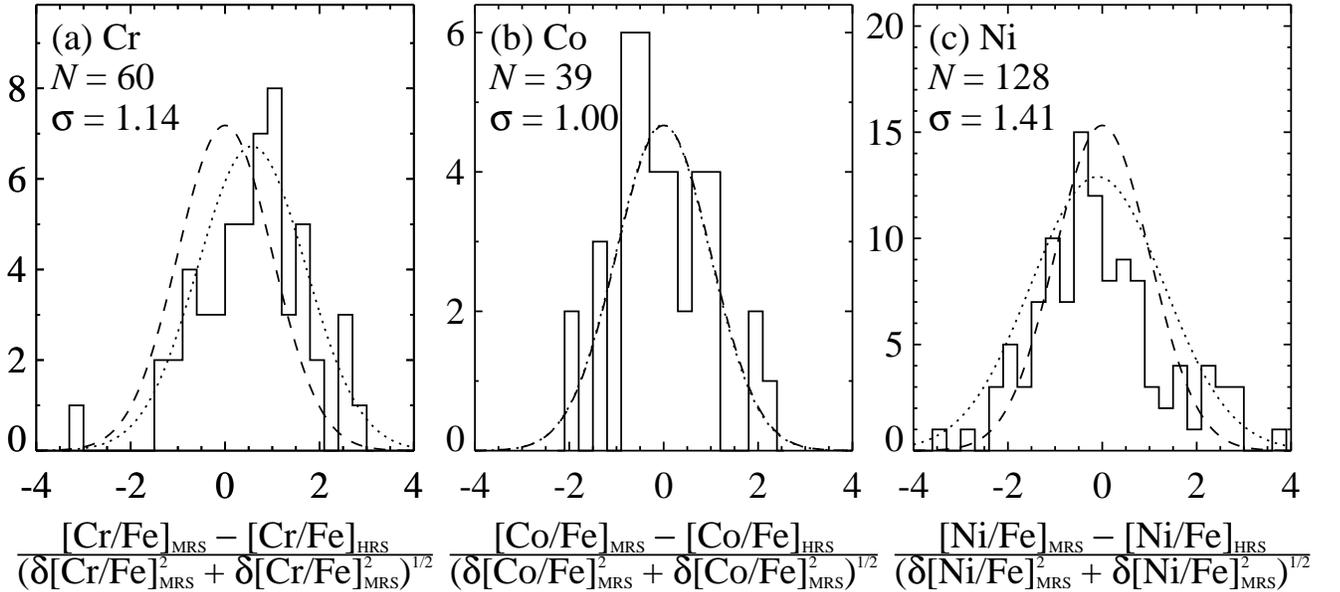}
\caption{The differences between our abundance measurements from MRS
  and literature values from HRS for the same stars.  The differences
  are normalized by the quadrature sum of the uncertainties.  If the
  errors are Gaussian and properly estimated, then these histograms
  would be Gaussians with a standard deviation of 1, as shown in
  Equation~\ref{eq:stddevdeltahrs}.  The dashed curves show Gaussians
  with $\sigma = 1$, whereas the dotted curves show the Gaussians that
  best represent the histograms.  Each panel gives the number of stars
  and actual standard deviations for each comparison
  sample.\label{fig:compare}}
\end{figure*}

We validated our medium-resolution spectroscopic (MRS) abundance
measurements by comparing them to high-resolution spectroscopic (HRS)
abundance measurements for the same stars, where available.  The
comparison sample largely overlaps with the HRS sample compiled from
the literature by K10 (see their Table~7).  Some stars from K10 are
not included either because we could not measure iron-peak abundances
from the DEIMOS spectra or because the literature source for the HRS
did not list those values.  We also supplemented the comparison set
with measurements from ultra-faint galaxies
\citep{sim07,kir08b,kir13b} and additional GCs (K16).  The comparison
sample includes \nhrs\ stars: \nhrsgc\ stars in GCs, \nhrshalo\ field
stars in the Milky Way's metal-poor halo, and \nhrsdsph\ stars in
dwarf galaxies.  We shifted all abundances onto our adopted solar
abundance scale.  None of the literature references applied NLTE
corrections to Cr or Co abundances.  We also use the element-to-iron
ratios rather than the absolute abundances (e.g., [Cr/Fe] rather than
$\epsilon({\rm Cr})$) because the ratio removes most of the effect of
differences in atmospheric parameters, especially $T_{\rm eff}$.
Also, the abundance ratio will be the more interesting quantity for
our future analyses in the context of galactic chemical evolution.

Table~\ref{tab:compare} gives the MRS and HRS measurements of
iron-peak abundances for the stars in common.  In cases of repeated
MRS measurements, the table includes multiple entries for the same
star.  For those entries, the HRS abundances are identical, but
different MRS measurements are given on different lines.
Figure~\ref{fig:validate} shows the differences between our
measurements and the HRS measurements from the literature.  The mean
differences ($\langle {\rm [X/Fe]}_{\rm MRS} - {\rm [X/Fe]}_{\rm HRS}
\rangle$) weighted by the quadrature sum of the HRS and MRS
uncertainties are $\crdiff \pm \crdifferr$ for Cr, $\codiff \pm
\codifferr$ for Co, and $\nidiff \pm \nidifferr$ for Ni.  The
differences are negligible for Co and Ni.  The difference for Cr is
small enough not to cause concern.  It could arise from a number of
factors, but we point out that the large majority of our DEIMOS
spectra include only three of the Cr absorption lines listed in
Table~\ref{tab:linelist}.  All three of these lines are red enough
that they are seldom included in HRS spectroscopic studies.  Thus, the
difference might be a result of using a different set of Cr lines.

If both the MRS and HRS measurements are accurate and their
uncertainties are properly estimated, then we expect the differences
normalized by the quadrature sum of their uncertainties to be normally
distributed with a mean of zero and standard deviation of unity.

\begin{eqnarray}
  \rm{mean}\left (\frac{{\rm [X/Fe]}_{{\rm MRS},i} - {\rm [X/Fe]}_{{\rm HRS},i}}{\sqrt{\delta{\rm [X/Fe]}_{\rm MRS}^2 + \delta{\rm [X/Fe]}_{\rm HRS}^2}} \right) = 0 \label{eq:meandeltahrs} \\
  \rm{stddev}\left (\frac{{\rm [X/Fe]}_{{\rm MRS},i} - {\rm [X/Fe]}_{{\rm HRS},i}}{\sqrt{\delta{\rm [X/Fe]}_{\rm MRS}^2 + \delta{\rm [X/Fe]}_{\rm HRS}^2}} \right) = 1 \label{eq:stddevdeltahrs}
\end{eqnarray}

\noindent
In fact, the means are $\mcr \pm \mcrerr$, $\mco \pm \mcoerr$, and
$\mni \pm \mnierr$ for Cr, Co, and Ni, respectively.  The standard
deviations are $\sdcr \pm \sdcrerr$, $\sdco \pm \sdcoerr$, and $\sdni
\pm \sdnierr$.  Figure~\ref{fig:compare} shows the distributions of
the quantity in parentheses in Equations~\ref{eq:meandeltahrs} and
\ref{eq:stddevdeltahrs}.  The distributions for Cr and Co do indeed
appear approximately Gaussian with standard deviations near 1\@.  The
distribution for Ni is positively skewed and leptokurtic, and its
formal standard deviation significantly exceeds 1.  These differences
indicate systematic differences between MRS and HRS in the measurement
of Ni and in the estimation of uncertainties.

Such differences are not surprising, given the large degree of
heterogeneity among the comparison sample, which is comprised of
\nhrsrefs\ different literature sources, most of which have different
line lists, abundance codes, and model atmospheres.  \citet{hin16}
showed that the results of even HRS spectroscopic analyses depend
sensitively on the choice of these various inputs.  (We also refer to
\citealt{sud17}\ for a comparison of different methods of determining
abundances in dwarf galaxies.)  Thus, the apparent imperfections shown
in Figure~\ref{fig:compare}(c) should not be taken as an indication of
serious problems in the abundance measurements.


\section{Summary}
\label{sec:summary}

The abundances of iron-peak elements trace the components of galactic
chemical evolution that are driven by explosive nucleosynthesis.
These abundances are especially sensitive to the explosion properties
of Type~Ia supernovae.  We have measured abundances of the iron-peak
elements Cr, Co, and Ni for \ntot\ metal-poor red giants in GCs, dwarf
galaxies, and the Milky Way halo.  We used mostly archival and some
new Keck/DEIMOS spectra, which we modeled with LTE synthetic spectra.
Table~\ref{tab:catalog} presents the catalog of measurements,
including previously measured abundances of Mg, Si, Ca, Ti, and Fe
(K10) and the new measurements of Cr, Co, and Ni.

We estimated systematic errors by enforcing that the abundance of each
element is constant within individual GCs.  The total uncertainty is
the quadrature sum of a fitting error that depends on S/N and the
systematic error.  We validated the appropriateness of the uncertainty
estimates by quantifying the dispersion of differences between repeat
measurements of the same stars.  The results indicated that the Cr and
Ni uncertainties are well-determined, but the Co uncertainties might
be slightly overestimated, at least at a given effective temperature.
We also validated the accuracy and precision of our measurements by
comparing them to high-resolution spectroscopic measurements of the
same stars.  The results are that our [Cr/Fe] measurements are
$(\crdiff \pm \crdifferr)$~dex higher than the high-resolution
literature values, and our [Ni/Fe] measurements are slightly more
discrepant with the literature values than indicated by our
uncertainty estimates.  Otherwise, the measurements agree within
expectations.

Our analysis does not reveal any serious concerns regarding the
accuracy of our measurements or our ability to assess the
uncertainties.  Thus, we have proven the capability to measure Cr, Co,
and Ni abundances from $R \sim 6500$ spectra in the wavelength range
6500--9000~\AA\@.  The major advantage of our data set is its
homogeneity.  The spectra were obtained on the same instrument in the
same configuration, and we measured the abundances with the same
software.  This represents the largest database of self-consistent
multi-element abundance measurements in both GCs and dwarf galaxies.

We attempted to apply NLTE corrections to our abundances, but we we
found that the corrections introduced trends in abundance with
effective temperature.  A better method of accounting for NLTE effects
is to compute a spectral grid in NLTE \citep[see][]{ruc13}.  These
calculations are underway, and we are planning to employ them in the
future.

Future work also includes measuring Mn, an iron-peak element that is
especially useful in diagnosing the explosion mechanism of Type~Ia
supernovae \citep{sei13}.  The wavelength range of the DEIMOS spectra
presented in this article does not include enough Mn lines for a
reliable measurement.  We intend to obtain bluer spectra in the future
to measure Mn abundances.

This article has not addressed any scientific interpretation of our
results.  An article in preparation will analyze the abundances in the
context of nucleosynthesis from various models of Type~Ia supernovae.

\acknowledgments 

We thank the anonymous referee for a constructive report.  This
material is based upon work supported by the National Science
Foundation under Grant No.\ AST-1614081.  Support for this work was
also provided by NASA through a grant from the Space Telescope Science
Institute (HST-GO-14734.011-A), which is operated by the Association
of Universities for Research in Astronomy, Inc., under NASA contract
NAS 5-26555.  ENK acknowledges funding from generous donors to the
California Institute of Technology.

The data presented herein were obtained at the W.~M.\ Keck
Observatory, which is operated as a scientific partnership among the
California Institute of Technology, the University of California and
the National Aeronautics and Space Administration.  The Observatory
was made possible by the generous financial support of the W.~M.\ Keck
Foundation.  The authors wish to recognize and acknowledge the very
significant cultural role and reverence that the summit of Maunakea
has always had within the indigenous Hawaiian community.  We are most
fortunate to have the opportunity to conduct observations from this
mountain.

\software{MOOG \citep{sne73}, MAFAGS-OS \citep{gru04a,gru04b}, MARCS
  \citep{gus08}, spec2d pipeline \citep{coo12,new13}, MPFIT
  \citep{mar12}}

\facility{Keck:II (DEIMOS)}

\bibliography{fepeak_catalog}
\bibliographystyle{apj}

\end{document}